\documentclass
[aps,prl,amsfonts,amssymb,twocolumn,amsmath,preprintnumbers,nofootinbib,floatfix,superscriptaddress,longbibliography]{revtex4-1}%
\usepackage[dvips]{graphics}
\usepackage{graphicx}
\usepackage{bm}
\usepackage{amsmath}
\usepackage{amsfonts}
\usepackage{amssymb}
\usepackage{xcolor}
\usepackage{subfigure}
\usepackage{tabularx}
\usepackage{multirow}
\usepackage{braket}
\usepackage{commath}
\usepackage[colorlinks,linkcolor=blue,anchorcolor=blue,citecolor=blue,urlcolor=blue]%
{hyperref}%
\makeatletter
\def\maketitle{
	\@author@finish
	\title@column\titleblock@produce
	\suppressfloats[t]}
\makeatother
\providecommand{\U}[1]{\protect\rule{.1in}{.1in}}

\begin{document}

\title{Crossed Nonlinear Dynamical Hall Effect in Twisted Bilayers }
\author{Cong Chen}
\thanks{These authors contributed equally to this work.}
\affiliation{New Cornerstone Science Laboratory, Department of Physics, University of Hong Kong, Hong Kong, China}
\affiliation{HKU-UCAS Joint Institute of Theoretical and Computational Physics at Hong Kong, Hong Kong, China}
\author{Dawei Zhai}
\thanks{These authors contributed equally to this work.}
\affiliation{New Cornerstone Science Laboratory, Department of Physics, University of Hong Kong, Hong Kong, China}
\affiliation{HKU-UCAS Joint Institute of Theoretical and Computational Physics at Hong Kong, Hong Kong, China}
\author{Cong Xiao}
\email{congxiao@um.edu.mo}
\affiliation{HKU-UCAS Joint Institute of Theoretical and Computational Physics at Hong Kong, Hong Kong, China}
\affiliation{Institute of Applied Physics and Materials Engineering, University of Macau, Taipa, Macau, China}
\author{Wang Yao}
\email{wangyao@hku.hk}
\affiliation{New Cornerstone Science Laboratory, Department of Physics, University of Hong Kong, Hong Kong, China}
\affiliation{HKU-UCAS Joint Institute of Theoretical and Computational Physics at Hong Kong, Hong Kong, China}

\begin{abstract}
We propose a novel nonlinear dynamical Hall effect characteristic of layered
materials with chiral symmetry, which is driven by the joint action of
in-plane and time variation of out-of-plane ac fields $\boldsymbol{j}%
_{\text{H}}\sim\boldsymbol{\dot{E}_{\perp}}\times\boldsymbol{E}_{\parallel}$.
A new band geometric quantity -- interlayer Berry connection polarizability,
which probes a mixed quantum metric characteristic of layer
hybridized electrons by twisted interlayer coupling, underlies this effect.
When the two orthogonal fields have common frequency, their phase difference
controls the on/off, direction and magnitude of the rectified Hall current. We
show sizable effects in twisted homobilayer transition metal dichalcogenides
and twisted bilayer graphene over broad range of twist angles.
Our work opens the door to discovering mixed quantum metric
responses unique to van der Waals stacking and concomitant applications under
the nonlinear spotlight.
\end{abstract}
\maketitle


Nonlinear Hall-type response to an in-plane electric field in a two
dimensional (2D) system with time reversal symmetry has attracted marked
interests \cite{Fu2015,Ma2019,Kang2019,Lu2021}. Intensive studies have been
devoted to uncovering new types of nonlinear Hall transport induced by quantum
geometry \cite{Lu2021,Lai2021,Zhou2022} and their applications such as
terahertz rectification \cite{Zhang2021} and magnetic information readout
\cite{Shao2020}. Restricted by symmetry \cite{Fu2015}, the known mechanisms of
nonlinear Hall response in quasi-2D nonmagnetic materials
\cite{Ma2019,Kang2019,He2022,Yao2022} are all of extrinsic nature, sensitive
to fine details of disorders \cite{Nagaosa2010,Du2019}, which have limited
their utilization for practical applications.

The intrinsic nonlinear Hall effect independent of scattering, on
the other hand, has been attracting increasing interest
\cite{Gao2014,Yan2020,Wang2021,Liu2021,Mazzola2023,Xiang2023}, and the very
recent observations of it in antiferromagnets spotlight the importance of
exploring Hall transport induced by quantum metric \cite{Xu2023QM,Gao2023QM}.
However, the intrinsic nonlinear Hall effect in its conventional paradigm
\cite{Gao2014} can only appear in magnetic materials.

Moreover, having a single driving field only, the conventional nonlinear Hall
effect has not unleashed the full potential of nonlinearity for enabling
controlled gate in logic operation, where separable inputs (i.e., in
orthogonal directions) are desirable. The latter, in the context of Hall
effect, calls for control by both out-of-plane and in-plane electric fields. A
strategy to introduce quantum geometric response to out-of-plane field in
quasi-2D geometry is made possible in van der Waals (vdW) layered structures
with twisted stacking
\cite{moireReviewEvaMacDonaldNatMater2020,moireReviewNatPhysBalents2020,moireReviewRubioNatPhys2021,moireReviewExptFolksNatRevMat2021,moireReviewJeanieLauNature2022,moireexcitonreviewNature2021,moireexcitonreviewNatRevMater2022}%
. Taking homobilayer as an example, electrons have an active layer degree of
freedom that is associated with an out-of-plane electric dipole
\cite{Guinea2012,Pesin2012,Xu2014}, whereas interlayer quantum tunneling rotates this
pseudospin about in-plane axes that are of topologically nontrivial textures
in the twisted landscapes \cite{WuMacDonaldPRL2019,HongyiNSR,Zhai2020PRL}.
Such layer pseudospin structures can underlie novel quantum geometric
properties when coupled with out-of-plane field.

In this work we unveil a new type of nonlinear Hall effect in time-reversal
symmetric twisted bilayers, where an intrinsic Hall current emerges under the
combined action of an in-plane electric field $\boldsymbol{E}_{\parallel}$ and
an out-of-plane ac field $\boldsymbol{E}_{\perp}(t)$: $\boldsymbol{j}%
\sim\boldsymbol{\dot{E}_{\perp}}\times\boldsymbol{E}_{\parallel}$ [see
Fig.~\ref{Fig:tTMD}(a)]. Having the two driving fields (inputs) and the
current response (output) all orthogonal to each other, the effect is dubbed
as the \textit{crossed nonlinear dynamical Hall effect}. This is also the
first nonlinear Hall contribution of an intrinsic nature in nonmagnetic
materials without external magnetic field, determined solely by the band
structures, not relying on extrinsic factors such as disorders and relaxation
times. Having two driving fields of the same frequency, a \textit{dc} Hall
current develops, whose on/off, direction and magnitude can all be controlled
by the phase difference of the two fields. The effect has a novel band
geometric origin in the momentum space curl of interlayer Berry connection
polarizability (BCP), {probes a mixed quantum metric} arising from the
interlayer hybridization of electronic states under the chiral crystal
symmetry, and enables a unique phase tunable rectification in chiral vdW
layered materials and a transport probe of them. As examples, we show sizable
effects in small angle twisted transition metal dichalcogenides (tTMDs) and
twisted bilayer graphene (tBG), as well as tBG of large angles where Umklapp
interlayer tunneling dominates.

\emph{{\color{blue} Geometric origin of the effect. }}A bilayer system couples
to in-plane and out-of-plane driving electric fields in completely different
ways. The in-plane field couples to the 2D crystal momentum, leading to
Berry-phase effects in the 2D momentum space \cite{Xiao2010}. In comparison,
the out-of-plane field is coupled to the interlayer dipole moment $\hat{p}$ in
the form of $-E_{\perp}\hat{p}$, where $\hat{p}=ed_{0}\hat{\sigma}_{z}$ with
$\hat{\sigma}_{z}$ as the Pauli matrix in the layer index subspace and $d_{0}$
the interlayer distance.
When the system has a more than twofold rotational axis in the $z$ direction,
as in tBG and tTMDs, any in-plane current driven by the out-of-plane field
alone is forbidden. It also prohibits the off-diagonal components of the
symmetric part of the conductivity tensor $\sigma_{ab}=\partial j_{a}/\partial
E_{||,b}$ with respect to the in-plane input and output. Since the
antisymmetric part of $\sigma_{ab}$ is not allowed by the Onsager reciprocity
in nonmagnetic systems, all the off-diagonal components of $\sigma_{ab}$ are
forbidden, irrespective of the order of out-of-plane field. On the other hand,
as we will show, an in-plane Hall conductivity $\sigma_{xy}=-\sigma_{yx}$ can
still be driven by the product of an in-plane field and the time variation
rate of an out-of-plane ac field.

To account for the effect, we make use of the semiclassical theory
\cite{Xiao2010,Gao2014,Xiao2021OM,Xiao2022MBT}. The velocity of an electron is
given by
\begin{equation}
\boldsymbol{\dot{r}}=\frac{1}{\hbar}\partial_{\boldsymbol{k}}\tilde
{\varepsilon}-\frac{e}{\hbar}\boldsymbol{E}_{\parallel}\times
\boldsymbol{\Omega}_{\boldsymbol{k}}-\boldsymbol{\Omega}_{\boldsymbol{k}%
E_{\perp}}\dot{E}_{\perp},
\end{equation}
with $\hbar\boldsymbol{k}$ as the 2D crystal momentum. Here and hereafter we
suppress the band index for simplicity, unless otherwise noted. For the
velocity at the order of interest, the $k$-space Berry curvature
$\boldsymbol{\Omega}_{\boldsymbol{k}}$ is corrected to the first order of the
variation rate of out-of-plane field $\dot{E}_{\perp}$:
\begin{equation}
\boldsymbol{\Omega}_{\boldsymbol{k}}=\partial_{\boldsymbol{k}}\times
(\boldsymbol{\mathcal{A}}+\boldsymbol{\mathcal{A}}^{\dot{E}_{\perp}}).
\end{equation}
Here $\boldsymbol{\mathcal{A}}=\langle u_{\boldsymbol{k}}|i\partial
_{\boldsymbol{k}}|u_{\boldsymbol{k}}\rangle$ is the unperturbed $k$%
-space\ Berry connection, with $|u_{\boldsymbol{k}}\rangle$ being the
cell-periodic part of the Bloch wave, whereas%
\begin{equation}
\boldsymbol{\mathcal{A}}^{\dot{E}_{\perp}}\left(  \boldsymbol{k}\right)
=\boldsymbol{\mathcal{G}}\left(  \boldsymbol{k}\right)  \dot{E}_{\perp}
\label{k-space}%
\end{equation}
is its gauge invariant correction \cite{Thouless1983,Dimi2006,Xiao2010}, which
can be identified physically as an in-plane positional shift of an electron
\cite{Gao2014} induced by the time evolution of the out-of-plane field. For a
band with index $n$, we have (details in the Supplemental Material
\cite{supp})
\begin{equation}
\boldsymbol{\mathcal{G}}^{n}\left(  \boldsymbol{k}\right)  =2\hbar
^{2}\mathrm{{\operatorname{Re}}}\sum_{m\neq n}\frac{p^{nm}\left(
\boldsymbol{k}\right)  \boldsymbol{v}^{mn}\left(  \boldsymbol{k}\right)
}{[\varepsilon_{n}\left(  \boldsymbol{k}\right)  -\varepsilon_{m}\left(
\boldsymbol{k}\right)  ]^{3}}, \label{BCP}%
\end{equation}
whose numerator involves the interband matrix elements of the interlayer
dipole and velocity operators, and $\varepsilon_{n}$ is the unperturbed band energy.

Meanwhile, up to the first order of in-plane field, the hybrid Berry curvature
in $\left(  \boldsymbol{k},E_{\perp}\right)  $ space reads $\boldsymbol{\Omega
}_{\boldsymbol{k}E_{\perp}}=\partial_{\boldsymbol{k}}(\mathfrak{\bm
A}+\mathfrak{\bm A}^{E_{||}})-\partial_{E_{\perp}}(\boldsymbol{\mathcal{A}%
}+\boldsymbol{\mathcal{A}}^{E_{||}})$. Here $\boldsymbol{\mathcal{A}}^{E_{||}%
}$ is the $k$-space Berry connection induced by $E_{||}$ field
\cite{Gao2014,Xiao2022MBT}, which represents an intralayer positional shift
and whose detailed expression is not needed for our purpose. $\mathfrak{\bm
A}=\langle u_{\boldsymbol{k}}|i\partial_{E_{\perp}}|u_{\boldsymbol{k}}\rangle$
is the $E_{\perp}$-space Berry connection \cite{Gao2020}, and
\begin{equation}
\mathfrak{\bm A}^{E_{||}}\left(  \boldsymbol{k}\right)  =\frac{e}{\hbar
}\boldsymbol{\mathcal{G}}\left(  \boldsymbol{k}\right)  \cdot\boldsymbol{E}%
_{\parallel} \label{lamda-space}%
\end{equation}
is its first order correction induced by the in-plane field. In addition,
$\tilde{\varepsilon}=\varepsilon+\delta\varepsilon$, where $\delta
\varepsilon=e\boldsymbol{E}_{\parallel}\cdot\boldsymbol{\mathcal{G}}\dot
{E}_{\perp}$ is the field-induced electron energy \cite{Xiao2021OM}.

Given that $\mathfrak{\bm A}^{E_{||}}$ is the $E_{\perp}$-space counterpart of
intralayer shift $\boldsymbol{\mathcal{A}}^{E_{||}}$, and that $E_{\perp}$ is
conjugate to the interlayer dipole moment, we can pictorially interpret
$\mathfrak{\bm A}^{E_{||}}$ as the interlayer shift induced by in-plane field.
It indeed has the desired property of flipping sign under the horizontal
mirror-plane reflection, hence is analogous to the so-called interlayer
coordinate shift introduced in the study of layer circular photogalvanic
effect \cite{Gao2020}, which is nothing but the $E_{\perp}$-space counterpart
of the shift vector well known in the nonlinear optical phenomenon of shift
current. Therefore,
the $E_{\perp}$-space BCP $e\boldsymbol{\mathcal{G}}/\hbar$ can be understood
as the interlayer BCP. This picture is further augmented by the connotation
that the interlayer BCP is featured exclusively by interlayer-hybridized
electronic states: According to Eq.~(\ref{BCP}), if the state $|u_{n}\rangle$
is fully polarized in a specific layer around some momentum $\boldsymbol{k}$,
then $\boldsymbol{\mathcal{G}}\left(  \boldsymbol{k}\right)  $ is suppressed.

With the velocity of individual electrons, the charge current density
contributed by the electron system can be obtained from $\boldsymbol{j}%
=e\int[d\boldsymbol{k}]f_{0}\boldsymbol{\dot{r}}$, where $[d\boldsymbol{k}]$
is shorthand for $\sum_{n}d^{2}\boldsymbol{k}/(2\pi)^{2}$, and the
distribution function is taken to be the Fermi function $f_{0}$ as we focus on
the intrinsic response. The band geometric contributions to $\boldsymbol{\dot
{r}}$ lead to a Hall current
\begin{equation}
\boldsymbol{j}=\chi^{\text{int}}\boldsymbol{\dot{E}_{\perp}}\times
\boldsymbol{E}_{\parallel}, \label{Hall}%
\end{equation}
where
\begin{equation}
\chi^{\text{int}}=\frac{e^{2}}{\hbar}\int[d\boldsymbol{k}]f_{0}[\partial
_{\boldsymbol{k}}\times\boldsymbol{\mathcal{G}}\left(  \boldsymbol{k}\right)
]_{z} \label{Eq:vorticity}%
\end{equation}
is intrinsic to the band structure. This band geometric quantity measures the
$k$-space curl of the interlayer BCP over the occupied states, and hence is
also a characteristic of layer-hybridized electronic states. Via an
integration by parts, it becomes clear that $\chi^{\text{int}}$ is a Fermi
surface property. Since $\chi^{\text{int}}$ is a time-reversal even
pseudoscalar, it is invariant under rotation, but flips sign under space
inversion, mirror reflection and rotoreflection symmetries. As such,
$\chi^{\text{int}}$ is allowed if and only if the system possesses a chiral
crystal structure, which is the very case of twisted
bilayers~\cite{Gao2020,ZhaiLayerHall2022}. Moreover, since twisted structures
with opposite twist angles are mirror images of each other, whereas the mirror
reflection flips the sign of $\chi^{\text{int}}$, the direction of Hall
current can be reversed by reversing twist direction.

\begin{figure*}[t]
\setlength{\abovecaptionskip}{0.cm}
\includegraphics[width=14 cm]{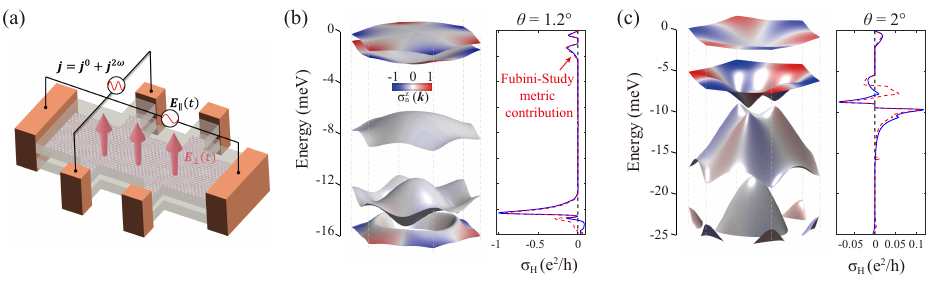}\caption{(a) Schematics of
experimental setup. (b, c) Valence band structure and intrinsic Hall
conductivity with respect to in-plane input for tMoTe$_{2}$ at twist angles
(b) $\theta=1.2^{\circ}$ and (c) $\theta=2^{\circ}$ in +K valley. In (b) and
(c) the color coding denotes the layer composition $\sigma_{n}^{z}%
(\boldsymbol{k})$, and the red dashed curve denotes the contribution of Fubini-Study metric term.}%
\label{Fig:tTMD}%
\end{figure*}

\emph{{\color{blue} Quantum metric nature of the effect. }}Given
the recent intensive studies on nonlinear Hall transport induced by
\textit{k}-space quantum metric \cite{Watanabe2021,Xu2023QM,Gao2023QM}, it is
interesting to point out that our proposed effect is ultimately related to the
mixed quantum metric in $\left(  \boldsymbol{k},E_{\perp}\right)  $ space,
which is unique to 2D layered materials. The interlayer BCP can be cast into
$\boldsymbol{\mathcal{G}}^{n}=-2\hbar\sum_{m\neq n}\boldsymbol{g}%
^{nm}/(\varepsilon_{n}-\varepsilon_{m})$, where $\boldsymbol{g}^{nm}%
=\mathrm{{\operatorname{Re}}}[\langle\partial_{E_{\perp}}u_{n}|u_{m}%
\rangle\langle u_{m}|\partial_{\boldsymbol{k}}u_{n}\rangle]$ has the meaning
of the quantum metric in $\left(  \boldsymbol{k},E_{\perp}\right)  $ space for
a pair of bands $n$ and $m$, in parallel to the familiar \textit{k}-space
quantum metric for a pair of bands \cite{Watanabe2021,Ahn2022,Bhalla2022}
(details in \cite{supp}). It is gauge invariant and related to the
Fubini-Study metric \cite{QM1980} $\boldsymbol{g}^{n}%
=\mathrm{{\operatorname{Re}}}\langle\partial_{E_{\perp}}u_{n}|(1-|u_{n}%
\rangle\langle u_{n}|)|\partial_{\boldsymbol{k}}u_{n}\rangle$ in $\left(
\boldsymbol{k},E_{\perp}\right)  $ space as $\boldsymbol{g}^{n}=\sum_{m\neq
n}\boldsymbol{g}^{nm}$. Moreover, $\chi^{\text{int}}$ can be decomposed into
the Fubini-Study metric term plus additional interband contribution (AIC)
\begin{equation}
\chi^{\text{int}}=2\hbar e^{2}\sum_{n}\int\frac{d\boldsymbol{k}}{\left(
2\pi\right)  ^{2}}\frac{\partial f_{0}}{\partial\varepsilon_{n}}%
\frac{(\boldsymbol{v}^{n}\times\boldsymbol{g}^{n})_{z}}{\varepsilon
_{n}-\varepsilon_{\bar{n}}}+\chi_{\text{AIC}}^{\text{int}},
\end{equation}
where $\boldsymbol{v}^{n}${$=\partial\varepsilon_{n}/\hbar\partial$%
}$\boldsymbol{k}$, and $\bar{n}$ denotes the band whose energy is closest to
$n$. In both tTMD and TBG, we find that the Fubini-Study metric term strongly
dominates (shown below in Figs.~\ref{Fig:tTMD}(b, c) and Figs.~\ref{Fig:tBG21}%
(e, f)).

\emph{{\color{blue} Phase tunable Hall rectification. }}This effect can be
utilized for the rectification and frequency doubling of an in-plane ac input
$\boldsymbol{E}_{\parallel}=\boldsymbol{E}_{\parallel}^{0}\cos\omega t$,
provided that the out-of-plane field has the same frequency, namely $E_{\perp
}=E_{\perp}^{0}\cos\left(  \omega t+\varphi\right)  $. The phase difference
$\varphi$ between the two fields plays an important role in determining the
Hall current, which takes the form of
\begin{equation}
\boldsymbol{j}=\boldsymbol{j}^{0}\sin\varphi+\boldsymbol{j}^{2\omega}%
\sin(2\omega t+\varphi).
\end{equation}
Here $\omega$ is required to be below the threshold for direct interband
transition in order to validate the semiclassical treatment, and
\begin{equation}
\boldsymbol{j}^{0}=\boldsymbol{j}^{2\omega}=\sigma_{\text{H}}\boldsymbol{\hat
{z}}\times\boldsymbol{E}_{\parallel}^{0},
\end{equation}
where $\sigma_{\text{H}}=\frac{1}{2}\omega E_{\perp}^{0}\chi^{\text{int}}$
quantifies the Hall
response with respect to the in-plane input.

One notes that the rectified output is allowed only if the two crossed driving fields are not
in-phase or anti-phase. Its on/off, chirality (right or left), and magnitude
are all controlled by the phase difference of the two fields. Such a unique
tunability provides not only a prominent experimental hallmark of this effect,
but also a controllable route to Hall rectification. In addition, reversing
the direction of the out-of-plane field switches that of the Hall current,
which also serves as a control knob.


\emph{{\color{blue} Application to tTMDs.}} We now study the effect
quantitatively in tTMDs, using tMoTe$_{2}$ as an
example~\cite{WuMacDonaldPRL2019,HongyiNSR} (see details of the continuum
model in \cite{ZhaiLayerHall2022}). For illustrative purposes, we take
$\omega/2\pi=0.1$~THz and $E_{\perp}^{0}d_{0}=10$~mV
\cite{moireReviewEvaMacDonaldNatMater2020,moireexcitonreviewNature2021,moireexcitonreviewNatRevMater2022}
in what follows.

\begin{figure}[t]
\includegraphics[width=3.0 in]{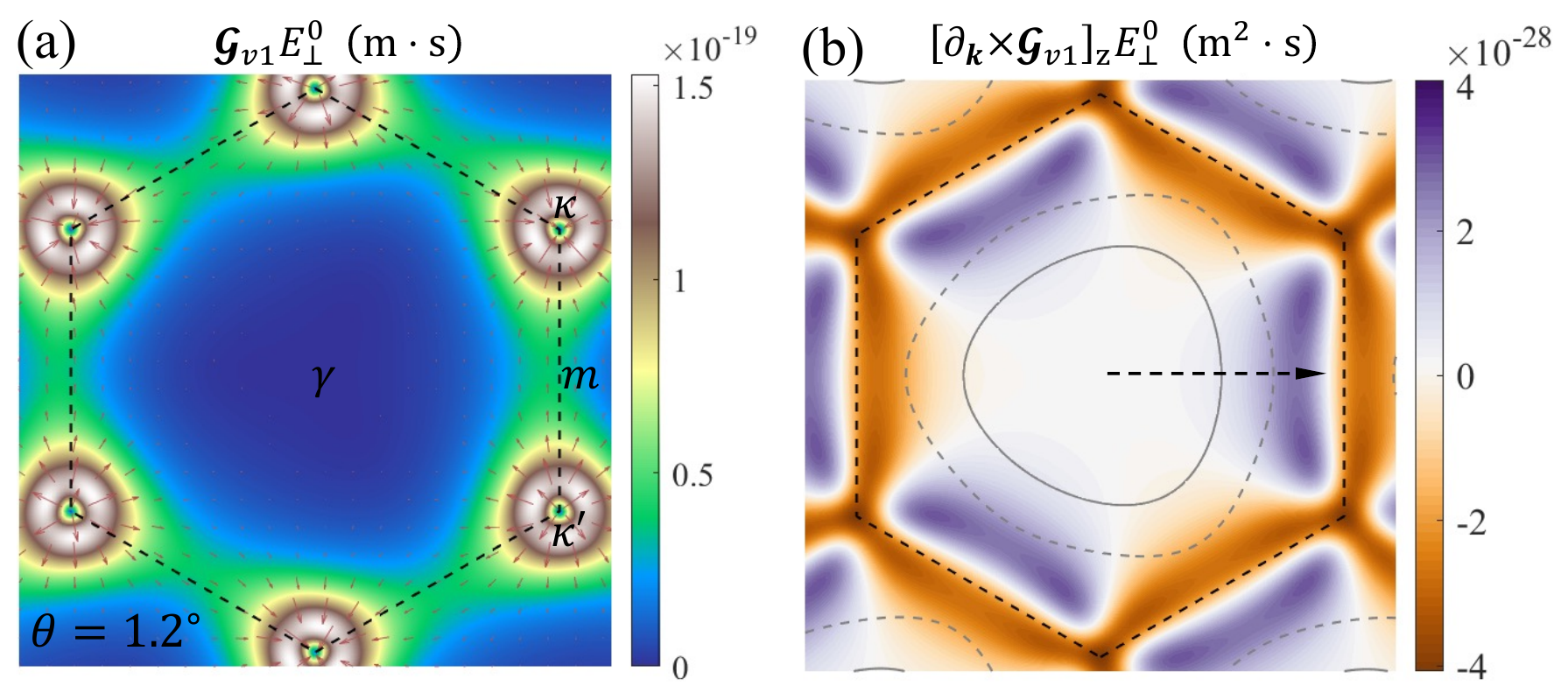} \caption{(a) The interlayer
BCP $\boldsymbol{\mathcal{G}}$, and (b) its vorticity $[\partial
_{\boldsymbol{k}}\times\boldsymbol{\mathcal{G}}]_{z}$ on the first valence
band from +K valley of 1.2$^{\circ}$ tMoTe$_{2}$. Background color and arrows
in (a) denote the magnitude and vector flow, respectively. Grey curves in (b)
show energy contours at $1/2$ and $3/4$ of the band width. The black dashed
arrow denotes direction of increasing hole doping level. Black dashed hexagons
in (a, b) denote the boundary of moir\'{e} Brillouin zone (mBZ).}%
\label{Fig:tTMD_distribution}%
\end{figure}

\begin{figure}[t]
\includegraphics[width=3.5 in]{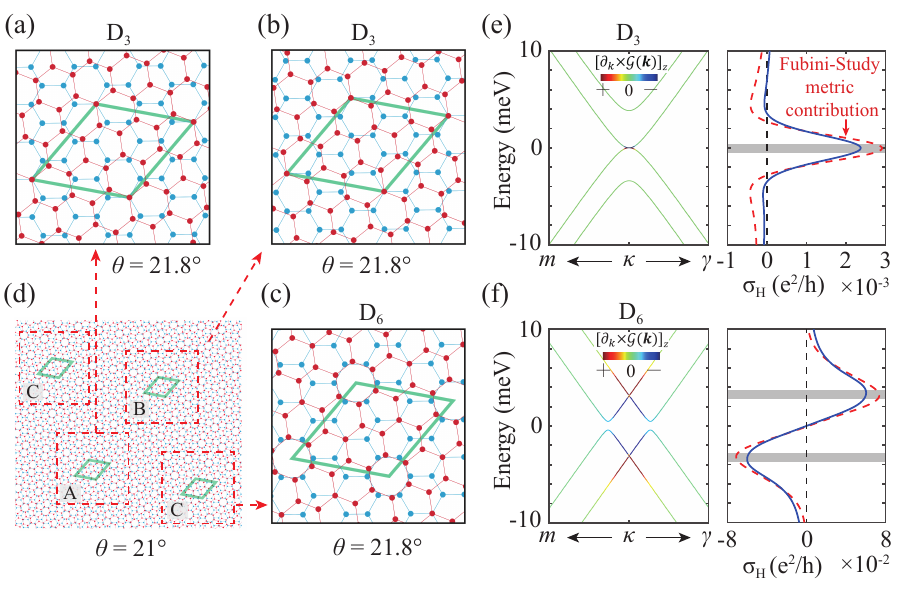}\caption{(a-c) Three high-symmetry
stacking registries for tBG with a commensurate twist angle $\theta
=21.8^{\circ}$. Lattice geometries with rotation center on an overlapping
atomic site (a, b) and hexagonal center (c). (d) Schematic of the moir\'{e}
pattern when the twist angle slightly deviates from $21.8^{\circ}$, here
$\theta=21^{\circ}.$ Red squares marked by A, B and C are the local regions
that resemble commensurate $21.8^{\circ}$ patterns in (a), (b) and (c),
respectively. (e, f) Low-energy band structures and intrinsic Hall
conductivity of the two geometries [(a) and (b) are equivalent].
The red dashed curve denotes the contribution of Fubini-Study
metric term. The shaded areas highlight energy windows $\sim\hbar\omega$
around band degeneracies where interband transitions, not considered here, may
quantitatively affect the conductivity measured.}%
\label{Fig:tBG21}%
\end{figure}

Figures~\ref{Fig:tTMD}(b) and (c) present the electronic band structures at twist angles
$\theta=1.2^{\circ}$ and $\theta=2^{\circ}$. In both cases, the energy spectra
exhibit isolated narrow bands with strong layer hybridization. At
$\theta=1.2^{\circ}$, the conductivity shows two peaks $\sim0.1e^{2}/h$ at low
energies associated with the first two valence bands. At higher hole-doping levels, a
remarkable conductivity peak $\sim e^{2}/h$ appears near the gap separating
the fourth and fifth bands. At $\theta=2^{\circ}$, the conductivity shows
smaller values, but the overall trends are similar: A peak $\sim
\mathcal{O}(0.01)e^{2}/h$ appears at low energies, while larger responses
$\sim\mathcal{O}(0.1)e^{2}/h$ can be spotted as the Fermi level decreases.

One can understand the behaviors of $\sigma_{\text{H}}$ from the interlayer
BCP in Eq.~(\ref{BCP}). It favors band near-degeneracy regions in
\textit{k}-space made up of strongly layer hybridized electronic states. As
such, the conductivity is most pronounced when the Fermi level is located
around such regions, which directly accounts for the peaks of response in
Fig.~\ref{Fig:tTMD}(b) [and \ref{Fig:tTMD}(c)]. When the Fermi level is
located on the third valence band in Fig.~\ref{Fig:tTMD}(b), the effect is
vanishingly small due to the large gaps to adjacent bands.

Let us take the case of Fermi level being located within the first valence
band of 1.2$^{\circ}$ tMoTe$_{2}$ in Fig.~\ref{Fig:tTMD}(b) as an example and explain the emergence of the
first conductivity peak. The \textit{k}-space distributions of
$\boldsymbol{\mathcal{G}}$ and $[\partial_{\boldsymbol{k}}\times
\boldsymbol{\mathcal{G}}]_{z}$ for this band are shown in Figs.~\ref{Fig:tTMD_distribution}(a) and
\ref{Fig:tTMD_distribution}(b), respectively. $\boldsymbol{\mathcal{G}}$ is
suppressed around the corners of mBZ, for the states are strongly layer
polarized there. Interlayer hybridization becomes stronger as $\boldsymbol{k}$
moves away from mBZ corners. In this process, the competition between enlarged
$p^{nm}(\boldsymbol{k})$ and \textit{k}-space local gap renders narrow
ring-like structures enclosing the mBZ corners, in which
$\boldsymbol{\mathcal{G}}$ is prominent and points radially inward/outward
around $\kappa$/$\kappa^{\prime}$. The distribution of
$\boldsymbol{\mathcal{G}}$ dictates that of $[\partial_{\boldsymbol{k}}%
\times\boldsymbol{\mathcal{G}}]_{z}$. One observes that $[\partial
_{\boldsymbol{k}}\times\boldsymbol{\mathcal{G}}]_{z}$ is negligible at lower
energies, and it is dominated by positive values as the doping increases, thus
the conductivity rises initially. As the doping level is higher, regions
of $[\partial_{\boldsymbol{k}}\times\boldsymbol{\mathcal{G}}]_{z}<0$ start
to contribute, thus the conductivity decreases after reaching a maximum.

\emph{{\color{blue} Application to tBG.}} The second example is tBG.
We focus on commensurate twist angles in the large angle limit in the main
text~\cite{PRB2013_Moon}, which possess moir{\'{e}}-lattice assisted strong
interlayer tunneling via Umklapp processes~\cite{PRB_Mele2010}. This case is
appealing because the Umklapp interlayer tunneling is a manifestation of
discrete translational symmetry of moir{\'{e} }superlattice, which is
irrelevant at small twist angles and not captured by the continuum model but
plays important roles in physical contexts such as higher order topological
insulator~\cite{Park2019} and moir{\'{e}} excitons
\cite{HongyiPRL2015,Seyler2019,Deng2020}. The Umklapp tunneling is strongest
for the commensurate twist angles of $\theta=21.8^{\circ}$ and $\theta
=38.2^{\circ}$, whose corresponding periodic moir{\'{e}} superlattices have
the smallest lattice constant ($\sqrt{7}$ of the monolayer counterpart). Such
a small moir\'{e} scale implies that the exact crystalline symmetry, which
depends sensitively on fine details of rotation center, has critical influence
on low-energy response properties.

To capture the Umklapp tunneling, we employ the tight-binding
model~\cite{PRB2013_Moon}. Figures~\ref{Fig:tBG21}(a, b) and (c) show two
distinct commensurate structures of tBG at $\theta=21.8^{\circ}$ belonging to
chiral point groups $D_{3}$ and $D_{6}$, respectively. The atomic
configurations in Figs.~\ref{Fig:tBG21}(a, b) are equivalent, which are
constructed by twisting AA-stacked bilayer graphene around an overlapping atom
site, and that in Fig.~\ref{Fig:tBG21}(c) is obtained by rotating around a
hexagonal center.
Band structures of these two configurations are drastically different within a
low-energy window of $\sim10$ meV around the $\kappa$
point~\cite{PRB2013_Moon}.
Remarkably, despite large $\theta$, we still get $\sigma_{\text{H}}$
$\sim\mathcal{O}(0.001)\,e^{2}/h$ ($D_{3}$) and $\sim\mathcal{O}%
(0.1)\,e^{2}/h$ ($D_{6}$), which are comparable to those at small angles (cf.
Fig.~S1 in \cite{supp}). Such sizable responses can
be attributed to the strong interlayer coupling enabled by Umklapp
processes~\cite{ZhaiLayerHall2022,HongyiPRL2015,Seyler2019,Deng2020}.
The profiles of $\sigma_{\text{H}}$ can be
understood from the distribution of $\left[\partial_{\boldsymbol{k}}%
\times\boldsymbol{\mathcal{G}}\right]  _{z}$.

Figure~\ref{Fig:tBG21}(d) illustrates the atomic structure of tBG with a twist
angle slightly deviating from $\theta=21.8^{\circ}$, forming a supermoir\'{e}
pattern \cite{supp}. In short range, the local stacking geometries resemble the
commensurate configurations at $\theta=21.8^{\circ}$, while the stacking
registries at different locales differ by a translation. Similar to the
moir\'{e} landscapes in the small-angle limit, there also exist high-symmetry
locales: Regions A and B enclose the $D_{3}$ structure, and region C contains
the $D_{6}$ configuration. Position-dependent Hall response is therefore
expected in such a supermoir\'{e}. As the intrinsic Hall signal from the
$D_{6}$ configuration dominates [see Figs.~\ref{Fig:tBG21}(e) vs (f)], the net
response mimics that in Fig.~\ref{Fig:tBG21}(f).
As the twist angle deviates
more from $21.8^{\circ}$, both the scales of supermoir\'{e} and of
the $D_3$ and $D_6$ local regions become shorter.

\emph{{\color{blue} Discussion. }}We have uncovered the intrinsic crossed
nonlinear dynamical Hall effect characteristic of layer hybridized electrons
in twisted bilayers, elucidated its quantum geometric origin, and
showed its sizable values in tTMD and tBG. Our focus is on the intrinsic
effect, which can be evaluated quantitatively for each material and provides a
benchmark for experiments. There may also be extrinsic contributions, similar
to the side jump and skew scattering in anomalous Hall effect. They
typically have distinct scaling behavior with the relaxation time $\tau$ from
the intrinsic effect, hence can be distinguished from the latter in
experiments \cite{Kang2019,Xiao2019scaling,Du2019,Lai2021}. Moreover, they are
suppressed in the clean limit $\omega\tau\gg1$ 
\cite{Du2019}. In high-quality tBG, $\tau\sim$ ps at room
temperature \cite{Sun2021}. Much longer $\tau$ can be obtained at lower
temperatures. In fact, a recent theory explaining well the resistivity of tBG
predicted $\tau\sim10^{-8}$ s at 10 K \cite{Sharma2021}. As such, high-quality
tBG under low temperatures and sub-terahertz input ($\omega/2\pi=0.1$ THz) is
located in the clean limit, rendering an ideal platform for isolating the
intrinsic effect.

This work paves a new route to driving in-plane response by out-of-plane
dynamical control of layered vdW structures \cite{zhai2022ultrafast}. The
study can be generalized to other observables such as spin current and spin
polarization, and the in-plane driving can be statistical forces, like
temperature gradient. Such orthogonal controls rely critically on the
nonconservation of layer pseudospin degree of freedom, and constitute an
emerging research field at the crossing of vdW materials, layertronics,
twistronics and nonlinear electronics.

This work is supported by the National Key R\&D Program of China (Grant No. 2020YFA0309600), the Research Grant Council of Hong Kong (AoE/P-701/20, HKU SRFS2122-7S05), the Croucher Foundation, and New Cornerstone Science Foundation. 
C.X. also acknowledges support by the UM Start-up Grant (SRG2023-00033-IAPME).

\bibliography{Ref}


\clearpage

\title{Supplemental Material}

\maketitle
\onecolumngrid
\tableofcontents

\renewcommand{\theequation}{S\arabic{equation}} \setcounter{equation}{0}
\renewcommand{\thefigure}{S\arabic{figure}} \setcounter{figure}{0}
\renewcommand{\thetable}{S\arabic{table}} \setcounter{table}{0}

\section{I. Derivation of Eqs.~(3) -- (5) in the main text}

Equation (5) in the main text shows that the first order correction induced by
in-plane field to the $E_{\perp}$-space Berry connection $\mathfrak{\bm
	A}=\langle u_{\boldsymbol{k}}|i\partial_{E_{\perp}}|u_{\boldsymbol{k}}\rangle$
is
\begin{equation}
\mathfrak{\bm A}^{E_{||}}\left(  \boldsymbol{k}\right)  =\frac{e}{\hbar
}\boldsymbol{\mathcal{G}}\left(  \boldsymbol{k}\right)  \cdot\boldsymbol{E}%
_{\parallel}. \label{lamda-space}%
\end{equation}
This can be derived from the field induced correction to the wave function. At
the first order of in-plane field, one has \cite{Gao2014,Xiao2022MBT}%
\[
|\delta^{E_{||}}u_{n\boldsymbol{k}}\rangle=\sum_{m\neq n}\frac{-\hbar
	\boldsymbol{\mathcal{A}}_{mn}\cdot e\boldsymbol{E}_{\parallel}/\hbar
}{\varepsilon_{n}-\varepsilon_{m}}|u_{m\boldsymbol{k}}\rangle,
\]
where $\boldsymbol{\mathcal{A}}_{mn}=\langle u_{m\boldsymbol{k}}%
|i\partial_{\boldsymbol{k}}|u_{n\boldsymbol{k}}\rangle$ is the \textit{k}%
-space interband Berry connection. We thus have%
\begin{align*}
\mathfrak{\bm A}^{E_{||}}\left(  \boldsymbol{k}\right)   &
=2\mathrm{{\operatorname{Re}}}\langle u_{n\boldsymbol{k}}|i\partial_{E_{\perp
}}|\delta^{E_{||}}u_{n\boldsymbol{k}}\rangle=-2e\sum_{m\neq n}\frac
{\mathrm{{\operatorname{Re}}}\left[  \langle u_{n}|i\partial_{E_{\perp}}%
	|u_{m}\rangle\langle u_{m}|i\partial_{\boldsymbol{k}}|u_{n}\rangle\right]
}{\varepsilon_{n}-\varepsilon_{m}}\cdot\boldsymbol{E}_{\parallel}\\
&  =2e\hbar\mathrm{{\operatorname{Re}}}\sum_{m\neq n}\frac{p^{nm}%
	\boldsymbol{v}^{mn}}{(\varepsilon_{n}-\varepsilon_{m})^{3}}\cdot
\boldsymbol{E}_{\parallel}\\
&  =\frac{e}{\hbar}\boldsymbol{\mathcal{G}}\left(  \boldsymbol{k}\right)
\cdot\boldsymbol{E}_{\parallel},
\end{align*}
with%
\begin{equation}
\boldsymbol{\mathcal{G}}^{n}=2\hbar^{2}\mathrm{{\operatorname{Re}}}\sum_{m\neq
	n}\frac{p^{nm}\boldsymbol{v}^{mn}}{(\varepsilon_{n}-\varepsilon_{m})^{3}%
}=-2\hbar\sum_{m\neq n}\frac{\mathrm{{\operatorname{Re}}}\left[  \langle
	u_{n}|i\partial_{E_{\perp}}|u_{m}\rangle\langle u_{m}|i\partial
	_{\boldsymbol{k}}|u_{n}\rangle\right]  }{\varepsilon_{n}-\varepsilon_{m}},
\label{BCP}%
\end{equation}
which proves Eqs. (4) and (5) in the main text.

Equation (3) in the main text shows that the first order correction induced by
the time-variation rate of out-of-plane field to the $k$-space Berry
connection $\boldsymbol{\mathcal{A}}=\langle u_{\boldsymbol{k}}|i\partial
_{\boldsymbol{k}}|u_{\boldsymbol{k}}\rangle$ is
\begin{equation}
\boldsymbol{\mathcal{A}}^{\dot{E}_{\perp}}\left(  \boldsymbol{k}\right)
=\boldsymbol{\mathcal{G}}\left(  \boldsymbol{k}\right)  \dot{E}_{\perp}
\label{k-space}%
\end{equation}
This can be also derived from the field induced correction to the wave
function. At the first order of $\dot{E}_{\perp}$, one has \cite{Dimi2006}%
\begin{equation}
|\delta^{\dot{E}_{\perp}}u_{n\boldsymbol{k}}\rangle=\sum_{m\neq n}\frac
{-\hbar\mathfrak{\bm A}_{mn}\dot{E}_{\perp}}{\varepsilon_{n}-\varepsilon_{m}%
}|u_{m\boldsymbol{k}}\rangle,
\end{equation}
where $\mathfrak{\bm A}_{mn}=\langle u_{m\boldsymbol{k}}|i\partial_{E_{\perp}%
}|u_{n\boldsymbol{k}}\rangle$ is the $E_{\perp}$-space interband Berry
connection. We thus have%
\[
\boldsymbol{\mathcal{A}}^{\dot{E}_{\perp}}\left(  \boldsymbol{k}\right)
=2\mathrm{{\operatorname{Re}}}\langle u_{n\boldsymbol{k}}|i\partial
_{\boldsymbol{k}}|\delta^{\dot{E}_{\perp}}u_{n\boldsymbol{k}}\rangle
=-2\hbar\sum_{m\neq n}\mathrm{{\operatorname{Re}}}\frac
{\boldsymbol{\mathcal{A}}_{nm}\mathfrak{\bm A}_{mn}}{\varepsilon
	_{n}-\varepsilon_{m}}\dot{E}_{\perp}=\boldsymbol{\mathcal{G}}\left(
\boldsymbol{k}\right)  \dot{E}_{\perp},
\]
which proves Eq. (3) in the main text.

\bigskip

\section{II. Quantum metric nature of the effect}

The very recent experimental observations of intrinsic nonlinear Hall effect
in antiferromagnets induced by \textit{k}-space quantum metric
\cite{Xu2023QM,Gao2023QM} spotlight the importance of exploring nontrivial
response induced by quantum metric in various extended parameter space.
Differing from that effect which comes from the \textit{k}-space quantum
metric and requires time-reversal symmetry breaking, our proposed effect comes
from the $\left(  \boldsymbol{k},E_{\perp}\right)  $-space quantum metric and
can appear in nonmagnetic systems. In this section of the Supplemental
Material, we show the quantum metric nature of the intrinsic crossed nonlinear
dynamical Hall effect.

The response coefficient of intrinsic crossed nonlinear dynamical Hall effect is%
\begin{equation}
\chi^{\text{int}}=\frac{e^{2}}{\hbar}\int[d\boldsymbol{k}]f_{0}[\partial
_{\boldsymbol{k}}\times\boldsymbol{\mathcal{G}}]_{z}, \label{Eq:vorticity}%
\end{equation}
which measures the antisymmetrized $k$-space dipole of the interlayer BCP over
the occupied states. Via an integration by parts, it becomes%
\begin{equation}
\chi^{\text{int}}=-e^{2}\int[d\boldsymbol{k}]f_{0}^{\prime}(\boldsymbol{v}%
\times\boldsymbol{\mathcal{G}})_{z},
\end{equation}
where $f_{0}^{\prime}=\partial f_{0}/\partial\varepsilon_{n}$. The numerator
of $\boldsymbol{\mathcal{G}}$ (Eq. (\ref{BCP}))
\begin{equation}
\boldsymbol{g}^{nm}=\mathrm{{\operatorname{Re}}}\left[  \langle u_{n}%
|i\partial_{E_{\perp}}|u_{m}\rangle\langle u_{m}|i\partial_{\boldsymbol{k}%
}|u_{n}\rangle\right]  =\mathrm{{\operatorname{Re}}}\left[  \langle
\partial_{E_{\perp}}u_{n}|u_{m}\rangle\langle u_{m}|\partial_{\boldsymbol{k}%
}u_{n}\rangle\right]
\end{equation}
is the quantum metric in $\left(  \boldsymbol{k},E_{\perp}\right)  $ space for
a pair of bands $n$ and $m$, in parallel to the well-known $\boldsymbol{k}%
$-space quantum metric for a pair of bands $\tilde{g}_{ij}^{nm}%
=\mathrm{{\operatorname{Re}}}\left[  \langle\partial_{k_{i}}u_{n}|u_{m}%
\rangle\langle u_{m}|\partial_{k_{j}}u_{n}\rangle\right]  $
\cite{Watanabe2021,Ahn2022,Bhalla2022}. It is gauge invariant and is related
to the Fubini-Study quantum metric \cite{QM1980} in $\left(  \boldsymbol{k}%
,E_{\perp}\right)  $ space as
\begin{equation}
\boldsymbol{g}^{n}=\sum_{m\neq n}\boldsymbol{g}^{nm}%
=\mathrm{{\operatorname{Re}}}\langle\partial_{E_{\perp}}u_{n}|\left(
1-|u_{n}\rangle\langle u_{n}|\right)  |\partial_{\boldsymbol{k}}u_{n}\rangle.
\end{equation}
The interlayer BCP is thus connected to the mixed quantum metric in $\left(
\boldsymbol{k},E_{\perp}\right)  $ space:%
\begin{equation}
\boldsymbol{\mathcal{G}}^{n}=-2\hbar\sum_{m\neq n}\frac{\boldsymbol{g}^{nm}%
}{\varepsilon_{n}-\varepsilon_{m}}.
\end{equation}
It can be decomposed into the Fubini-Study metric (FSM) term and an additional
interband contribution (AIC)%
\begin{equation}
\boldsymbol{\mathcal{G}}^{n}=\boldsymbol{\mathcal{G}}_{\text{FSM}}%
^{n}+\boldsymbol{\mathcal{G}}_{\text{AIC}}^{n},
\end{equation}
where%
\begin{align}
\boldsymbol{\mathcal{G}}_{\text{FSM}}^{n}  &  \equiv-2\hbar\frac
{\boldsymbol{g}^{n}}{\varepsilon_{n}-\varepsilon_{\bar{n}}},\\
\boldsymbol{\mathcal{G}}_{\text{AIC}}^{n}  &  \equiv-2\hbar\sum_{m\neq
	n,\bar{n}}\boldsymbol{g}^{nm}\frac{\varepsilon_{m}-\varepsilon_{\bar{n}}%
}{\left(  \varepsilon_{n}-\varepsilon_{m}\right)  \left(  \varepsilon
	_{n}-\varepsilon_{\bar{n}}\right)  },
\end{align}
with $\bar{n}$ being the band whose energy is closest to $n$. Accordingly, the
response coefficient of intrinsic crossed nonlinear dynamical Hall effect,
\begin{equation}
\chi^{\text{int}}=2\hbar e^{2}\sum_{n}\sum_{m\neq n}\int\frac{d\boldsymbol{k}%
}{\left(  2\pi\right)  ^{2}}f_{0}^{\prime}\frac{(\boldsymbol{v}^{n}%
	\times\boldsymbol{g}^{nm})_{z}}{\varepsilon_{n}-\varepsilon_{m}},
\end{equation}
is related to the mixed quantum metric in $\left(  \boldsymbol{k},E_{\perp
}\right)  $ space, and can be docmposed into
\begin{equation}
\chi^{\text{int}}=\chi_{\text{FSM}}^{\text{int}}+\chi_{\text{AIC}}%
^{\text{int}},
\end{equation}
where the U(1) quantum metric term reads
\begin{align}
\chi_{\text{FSM}}^{\text{int}}  &  \equiv-e^{2}\sum_{n}\int\frac
{d\boldsymbol{k}}{\left(  2\pi\right)  ^{2}}f_{0}^{\prime}(\boldsymbol{v}%
^{n}\times\boldsymbol{\mathcal{G}}_{\text{QM}}^{n})_{z}\nonumber\\
&  =2\hbar e^{2}\sum_{n}\int\frac{d\boldsymbol{k}}{\left(  2\pi\right)  ^{2}%
}f_{0}^{\prime}\frac{(\boldsymbol{v}^{n}\times\boldsymbol{g}^{n})_{z}%
}{\varepsilon_{n}-\varepsilon_{\bar{n}}},
\end{align}
and the AIC is given by%
\begin{align}
\chi_{\text{AIC}}^{\text{int}}  &  \equiv-e^{2}\sum_{n}\int\frac
{d\boldsymbol{k}}{\left(  2\pi\right)  ^{2}}f_{0}^{\prime}(\boldsymbol{v}%
^{n}\times\boldsymbol{\mathcal{G}}_{\text{AIC}}^{n})_{z}\nonumber\\
&  =2\hbar e^{2}\sum_{n}\sum_{m\neq n,\bar{n}}\int\frac{d\boldsymbol{k}%
}{\left(  2\pi\right)  ^{2}}f_{0}^{\prime}(\boldsymbol{v}^{n}\times
\boldsymbol{g}^{nm})_{z}\frac{\varepsilon_{m}-\varepsilon_{\bar{n}}}{\left(
	\varepsilon_{n}-\varepsilon_{m}\right)  \left(  \varepsilon_{n}-\varepsilon
	_{\bar{n}}\right)  }.
\end{align}

\section{III. Extra figures for tBG at small twist angles}

\begin{figure}[th]
	\includegraphics[width=4 in]{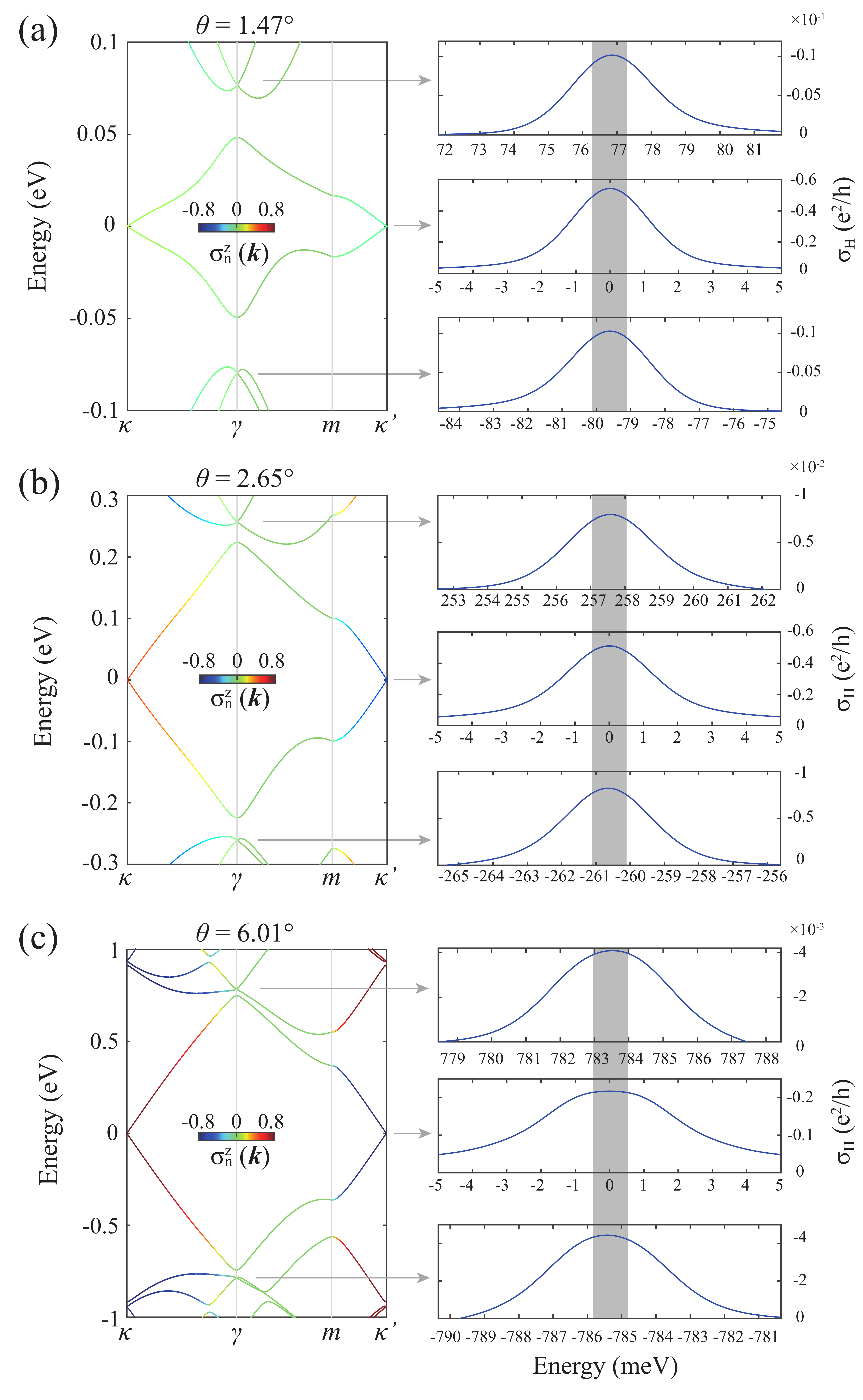}\caption{Band structure and layer
		composition $\sigma_{n}^{z}$ in +K valley of tBG (left panel) and the
		intrinsic Hall conductivity (right panel) at three different twist angle
		$\theta$. The shaded areas highlight energy windows $\sim\hbar\omega$ around
		band degeneracies in which the conductivity results should not be considered.
		Here $\sigma_{H}$ should be multiplied by a factor of 2 accounting for spin
		degeneracy.}%
	\label{Fig:tBG}%
\end{figure}

Figure~\ref{Fig:tBG}(a) shows the band structure of tBG with $\theta
=1.47^{\circ}$ obtained from the continuum model~\cite{KoshinoTBGPRX2018}. The
central bands are well separated from higher ones, and show Dirac points at
$\kappa$/$\kappa^{\prime}$ points protected by valley $U(1)$ symmetry and a
composite operation of twofold rotation and time reversal $C_{2z}\mathcal{T}%
$~\cite{PRL_2019_SongZD}. Degeneracies at higher energies can also be
identified, for example, around $\pm75$~meV at the $\gamma$ point. As the two
Dirac cones from the two layers intersect around the same area, such
degeneracies are usually accompanied by strong layer hybridization [see the
color in the left panel of Fig.~\ref{Fig:tBG}(a)]. Additionally, it is
well-known that the two layers are strongly coupled when $\theta$ is around
the magic angle ($\sim1.08^{\circ}$), rendering narrow bandwidths for the
central bands. As discussed in the main text, coexistence of strong interlayer
hybridization and small energy separations is expected to contribute sharp
conductivity peaks near band degeneracies, as shown in Fig.~\ref{Fig:tBG}(a).
In this case, the conductivity peak near the Dirac point can reach
$\sim0.1e^{2}/h$, while the responses around $\pm0.08$~eV are smaller at
$\sim0.01e^{2}/h$.

The above features are maintained when $\theta$ is enlarged, as illustrated in
Figs.~\ref{Fig:tBG}(b) and (c) using $\theta=2.65^{\circ}$ and $\theta
=6.01^{\circ}$. Since interlayer coupling becomes weaker and the bands are
more separated at low energies when $\theta$ increases, intensity of the
conductivity drops significantly.

We stress that $\boldsymbol{\mathcal{G}}$ is not defined at degenerate points,
and interband transitions may occur when energy separation satisfies
$|\varepsilon_{n}-\varepsilon_{m}|\sim\hbar\omega$, the effects of which are
not included in the current formulations. Consequently, the results around
band degeneracies within energy $\sim\hbar\omega$ [shaded areas in
Fig.~\ref{Fig:tBG}] should be excluded.

\clearpage

\section{IV. Detailed discussions on tBG with twist angles around $21.8^{\circ}$}

\begin{figure}[th]
	\includegraphics[width=6 in]{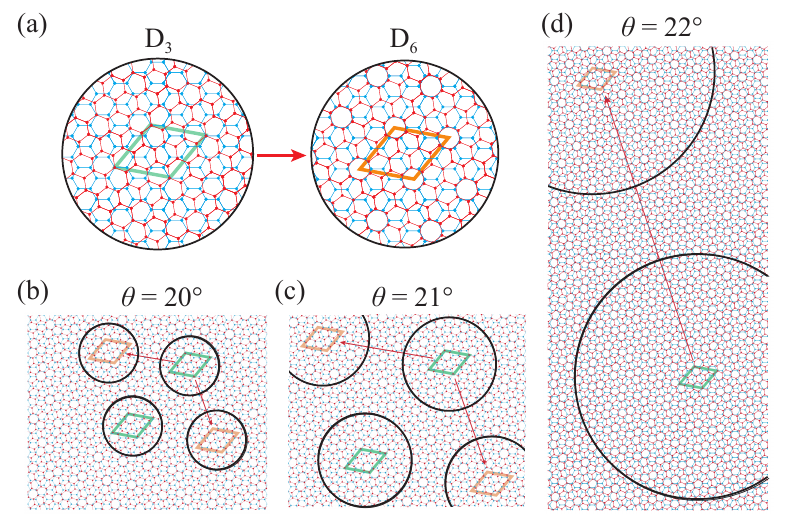}\caption{(a) Commensurate tBG
		structures at $\theta= 21.8^{\circ} $ with $D_{3}$ or $D_{6}$ symmetries.
		(b--d) Schematics of supermoir{\'{e}} pattern when twist angle deviates from
		$\theta= 21.8^{\circ}$. Local regions that resemble the $D_{3}$ and the $D_{6}$
		commensurate structures are enclosed by circles.}%
	\label{Fig:tBG_deviation}%
\end{figure}

Figure~\ref{Fig:tBG_deviation}(a) shows the two distinct commensurate
structures of tBG at $21.8^{\circ}$ belonging to chiral point group $D_{3}$
and $D_{6}$, respectively. Figures~\ref{Fig:tBG_deviation}(b--d) show lattice
structures with various derivations (from large to small) from the
commensurate angle $21.8^{\circ}$. Recall that in small-angle moir{\'{e}}
patterns, there exist high-symmetry local stacking regions that
resemble the commensurate AA, AB, or BA configurations.
Likewise, in the supermoir{\'{e}} pattern formed by deviating slightly from
$21.8^{\circ}$, there exist local regions enclosed by the circles in Figs.~\ref{Fig:tBG_deviation}(b--d) whose landscapes resemble the
commensurate $D_{6}$ and $D_{3}$ structures in (a).
The period of the supermoir{\'{e}} scales linearly with the distance between two adjacent circles. Along
the directions of the arrows in Figs.~\ref{Fig:tBG_deviation}(b--d), the local lattice structure smoothly alternates
between $D_{3}$ and $D_{6}$ stackings. When the deviation from $21.8^{\circ}$ is small, e.g., Figs.~\ref{Fig:tBG_deviation}(c, d), the
supermoir{\'{e}} contains many $D_{3}$ and $D_{6}$ unit cells. We expect that the
Hall conductivity is dominantly contributed by the $D_{6}$ stackings. As the
derivation gets larger, both the scales of supermoir\'{e} and of
local regions with commensurate structures become shorter.

\end{document}